\documentclass[journal]{IEEEtran}
\usepackage{blindtext}
\usepackage{graphicx}
\usepackage{amsmath}
\usepackage{amssymb}
\usepackage{listings}
\usepackage{color}
\usepackage{hyperref}
\usepackage{cite}

\begin{document}

\title{Electric Vehicle Charge Scheduling on Highway Networks from an Aggregate Cost Perspective}

\author{Sean Anderson and Vineet Nair}

\maketitle
\begin{abstract}

In this paper, we attempt to optimally schedule the charging of long-range battery electric vehicles (BEVs) along highway networks, in order to minimize aggregate costs to the overall system consisting of utilities or electricity providers, station operators and other infrastructure, as well as EV users. Thus, we approach the problem from the perspective of both customers (EV car owners), as well as charging station operators and utilities using a hybrid systems based formulation.

\end{abstract}

\vspace{-0.15in}
\section{Introduction}
\subsection{Motivation}

The popularity of electric vehicles has been rising rapidly in recent years, driven mainly by the greater driving comfort, improved safety, cheaper maintenance and lower exhaust emissions they offer. As battery and materials costs continue to decline and we transition towards cleaner, more diversified and renewable electricity generation, the use of both plug-in hybrid and battery electric vehicles is likely to expand even more. 

Increased EV utilization rates also mean that we will need more sophisticated and effective techniques to schedule the charging of both plug-in and battery electric vehicles in real-time, and allocate charging capacity or power in the most efficient manner. This is in order to minimize negative effects like congestion, long waiting times for customers and high infrastructure costs like uneven demands or loads for utilities and station operators. On a broader level, optimal EV charging will also play an important role in realizing smart grids of the future. 

\subsection{Literature Review}
The work proposed thus far for EV scheduling has been focused on four areas: optimization with respect to load on power distribution networks \cite{tangetal},\cite{matal},\cite{sortommeetal},\cite{ganetal},\cite{heetal2}, queuing models for input-output dynamics of charging stations \cite{saidetal},\cite{gusrialdietal},\cite{qinetal},\cite{tanetal}, game-based approaches for trajectory optimization \cite{gusrialdietal},\cite{mohammadietal},\cite{heetal}, and more traditional trajectory optimal control methods \cite{kongetal},\cite{eisneretal},\cite{schneideretal},\cite{bayrametal},\cite{jinetal}. Note that power distribution problems either use game-based approaches or traditional optimal control methods but are considered distinct in their objectives: the former looks at power quality on networks whereas the latter has objective functions related to time or expected financial costs.

The work done in power distribution considers the effect of EV charging on the grid, mainly with regard to base load. In this sense EVs are treated as dispatchable loads that can fill valleys during otherwise low demand. Other work has considered topological effects of charging station power draws. The work described in \cite{tangetal} uses statistical estimation to predict future EV arrivals and the associated charging demands in order to perform load-shifting via model predictive control. \cite{matal} proposes a mean-field game-based approach for large EV populations. The objective is to fill the overnight valley in base load and does not consider EV routing. The work in \cite{sortommeetal} coordinates plug-in hybrid EVs (PHEV) in order to reduce system power losses, maximize load factor, and minimize load variance. A decentralized approach is proposed in \cite{ganetal} that focuses on load shifting via dispatch signals provided by a centralized authority. The issue of load shifting is considered in \cite{heetal2} with emphasis on circumventing load forecasting limitations. A locally optimal distributed approach is demonstrated. The work mentioned here focuses on optimization in the time domain--load shifting for valley filling throughout a day.

Many different queuing model approaches have been proposed for optimal EV scheduling. These works generally focus on modeling the arrival and departure/service rates of the EVs. While some approaches utilize probability theory as the foundation for the scheduling, others rely on optimization. The work in \cite{saidetal} presents a first-come first-served queuing model with arrival rates corresponding to an exponential distribution and Poisson departure rates in a Markov chain. This does not account for location and path planning. The dynamics of the EV battery are not considered. A similar approach is shown in \cite{gusrialdietal} where the objective is to minimize EV waiting time at charging stations. This makes more considerations around path planning between nodes but limits the road network to a one-dimensional, unidirectional flow. \cite{qinetal} also minimizes EV waiting time using a similar queuing model. Theoretical bounds on waiting time are presented as well as a communication strategy (vehicle-to-charging station). A battery-swapping station approach in \cite{tanetal} utilizes queues and also defines probabilities around customer dissatisfaction.

Trajectory optimization problems have been formulated for electric vehicle charging with an emphasis on finding efficient algorithms and sensible objective functions. In \cite{kongetal} a hierarchical approach is proposed that considers power system network dynamics and battery dynamics. Given a fleet of EVs the highest level problem is station location. The provisioning problem sits below this in which the number of chargers per station is allocated according to a queuing model and the given demand. The final layer consists of a global cost minimization. The work in \cite{eisneretal} focuses on a graph theoretic framework for EV scheduling. The work exploits previous classical work done in path planning to reduce the complexity of scheduling EV charging in large networks. \cite{schneideretal} details heuristic search algorithms for solving the EV-routing problem on a road network with charging stations located at specific nodes. A mixed approach is described in \cite{bayrametal} whereby probabilistic tools, power distribution concerns, and an objective function formulated for routing efficiency are used. More specifically, the problem maximizes the station profits while maintaining a high equality of service. Finally, the work in \cite{jinetal} utilizes finite horizon and receding horizon approaches to consider the two cases of 1) a priori knowledge of all users and 2) new inputs (vehicles) to the system during the horizon for vehicle routing on a charging network. The objective function takes into consideration the customer's perspective in an attempt to satisfy demand while keeping the cost low. Here a variable charging rate is implemented, which adds complexity to the required infrastructure and solution space but enables more interesting solutions.

\section{Contributions}
In this paper, we propose looking at the scheduling problem for EV charging from the perspective of minimizing \textit{aggregate} costs to the system as a whole. This addresses an important null space in the literature since much of the previous work in this area has focused on optimizing the allocation of charging power with respect to only one of the players i.e. utilities, charging station operators, or customers. 

A lot of the past research has also focused on determining optimal locations for charging stations in a given area. However, in this paper, we take the station locations (i.e. the nodes) as a fixed parameter for a particular city or urban area's highway network. Instead, we aim to design a central controller or algorithm that determines the optimal trajectory that an EV can take given their starting position and desired destination, in a manner that maximizes benefits to all stakeholders involved.

Furthermore, a great deal of effort has gone into minimizing losses to utilities and the grid - and utilizing EVs to provide services like load flattening, voltage and frequency regulation. However, we place more emphasis on the opportunity costs paid by customers as well, in terms of both time spent and physical degradation of their car batteries. Following the prescribed route will also help alleviate range anxiety among users, and potentially contribute towards increased EV penetration.

Additionally, this work defines a hybrid systems theoretic model and uses finite state machines to describe the dynamics of electric vehicles on a highway network under two different frameworks. The model is first described under a framework that is more easily converted into code. The second description allows for more thorough analysis of the model's properties.

\section{Model} \label{sec: model}

The following model definitions and formulations formalize the Mixed Integer Quadratic Programming (MIQP) problem.

\subsection{Definitions and Notation}

Table \ref{tab: defs} defines the basic terms that are used throughout the work. The remainder of the notation is defined in-text when first introduced.

\begin{table}
    \begin{center}
    \begin{tabular}{ | p{4.25 cm} | p{3.75 cm} | }
    \hline
     \centering \textbf{Variable and Symbol} & \textbf{Notes}  \\
    \hline
    \centering Car index $c \in \{1,......,p\}$ & p = Total no. of cars \\ 
    \hline
     \centering Time step $k \in \{1,.....,H_p\}$ & $H_p$ = Prediction horizon \\  
    \hline
    \centering Station index $n \in \{1,....N\}$ & N = Total no. of stations \\
    \hline
    \centering $\nu_0$ & Starting node for a car\\
    \hline
     \centering $\nu_{H_p}$ & Ending node for a car \\
    \hline
    \centering $d_k$ & Discrete trip distance counter\\
    \hline
    \centering $\epsilon_k$ & Discrete local (edge) distance counter\\
    \hline
    \centering $\gamma \in \{0,1\}$ & Constrained input: at a node or on an edge\\
    \hline
    \centering $y \in \{0,1\}$ & Constrained input: charging or not\\
    \hline
    \centering $E_k$ & Energy level of an EV at time k \\
    \hline
    \end{tabular}
    \end{center}
    \caption{Definitions of general notation}
    \label{tab: defs}
\end{table}

\subsection{Key Assumptions of Formulation}

\begin{itemize}
    \item All nodes or intersections in our highway network (represented as a graph) have a charging station.
    \item All EVs in our network have identical battery energy level capacities and the same discharging power dynamics while driving.
    \item All charging stations in our network provide identical identical charging power profiles for all EV customers.
\end{itemize}

\subsection{Highway Network Model}
A highway network can be described as a connected graph $\mathcal{G} = (\mathcal{V},\mathcal{E},A)$ where $\mathcal{V} = \{1,..,N\}$ is a non-empty set of N nodes that represent road intersections and the potential location of a EV charging station, $\mathcal{E}$ is the set of edges representing the road segment lengths, and $A$ is the adjacency matrix where $a^{ii}= 0$ and $a^{ij} > 0$ for all connected nodes $i \neq j$. The adjacency matrix defines whether nodes are connected. For the set of edges, $\mathcal{E}$, an entry $e^{ij} > 0$ indicates the transition from node $i$ to node $j$ with a positive weight. Generally, a graph can be directed or undirected. Considering that the graph represents a highway, traffic is assumed to be able to flow both ways without loss of generality. This can be generalized to cases with only one-way traffic flows, because if $\exists$ $e^{ij} > 0 $ then and physically $\nexists$ $e_{ji} >  0$, then $e_{ji} := 0$ where the convention is taken to be that only positive-valued edges can be traversed. Equivalently, the zero-valued edges can be assigned to $NaN$ or any placeholder value. Finally, the neighborhood of a node is defined where $\mathcal{N}_{i} := \{j| (i,j) \in \mathcal{E} \text{ s.t. } e^{ij} > 0\}$.

The graph representation helps to formalize the work in notation but no novel or extensive graph theory is used here.

\begin{figure}[h!] 
  \includegraphics[scale=.12]{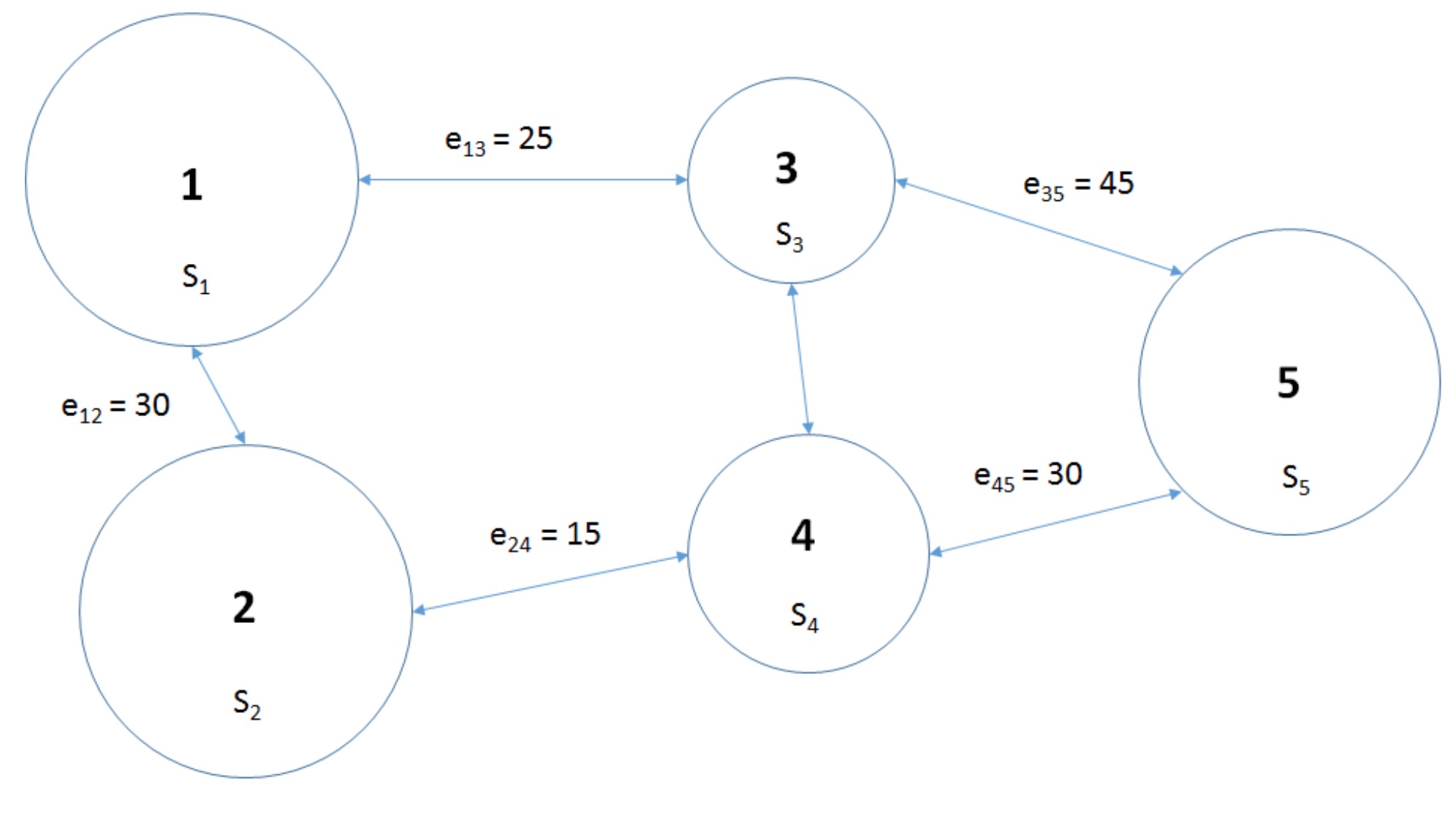}
  \caption{Nodes represent charging stations. Node sizes correspond to station capacities. Edge weights represent distances in miles. Assume all nodes connected through two-way highways/roads results in an undirected graph}
  \label{fig: network}
\end{figure}

The network used in the simulations is presented in figure \ref{fig: network}. It is included here to help motivate the application of graph theory to the EV network.

\subsection{Costs for the Network} \label{eq: network_cost}

\begin{align}
    C_{electricity} &= \sum_{n = 1}^N \sum_{k = 1}^{H_p-1} \sum_{c = 1}^p P_{k}^{elec}(n) * \; y_{k}^{c} * \;
    (E_{k+1}-E_{k})
\end{align}

Let $P^{elec}_{k}(n)$ represent the electricity price (dollars/kWh) at a particular station n, as a function of location in the network and according to the time step k. The difference in energy when it is charging (i.e. when $y_k$ is true) is taken to be the amount of electricity, in kWh, that was just injected into the battery. This cost term then is in terms of dollars.

We originally considered including electricity costs using Time-of-Use (TOU) rates as part of stations' costs, as described in equation \eqref{eq: network_cost} above. However, we later decided against this since previous studies \cite{heetal} have shown that electricity costs are much smaller than costs associated with travel times. Thus, we chose time minimization for customers as the dominant decision criterion, where travelers choose the fastest feasible path while also satisfying the other constraints.

\begin{align} 
    \begin{split}
        C_{station} &= \sum_{k = 1}^{H_{p}} \sum_{n = 1}^{N}
        \bar{C} \; * \mid sgn\bigg(\frac{U(k, n)}{S_{n}} - 0.5\bigg)\mid * \\ 
        & \qquad \; (U(k,n) - S_{n})  
    \end{split} \\
    U(k,n) &= \sum_{c=1}^{p} y_k^c(n) \label{eq: network_costs}
\end{align}

In order to avoid either excessive under or over-utilization of charging stations and ensure a uniform burden on all nodes across the network, we decided to penalize stations whenever they use above or below 50\% of their maximum rated capacity, as shown in \eqref{eq: network_costs}. We imposed a flat linear cost $\bar{C}$ on the difference between the current station utilization and the preferred capacity of that station $S_n$. For simplicity, we defined station utilization $U(k,n)$ in terms of the number of cars charging at a particular station $n$ at any given time $k$. This can be easily computed by summing over the boolean variables $y_k^c(n)$ in our logic FSM described in Section \ref{sec: logic_fsm}, as shown in \eqref{eq: network_costs} above. In general the following definition holds:

\begin{align}
    sgn\bigg(\frac{U(k, n)}{S_{n}} - 0.5\bigg) = 
    \begin{cases} 
    1, & U(k,n) > 50 \% \; of \; S_n \\
    0, & U(k,n) = S_n \\
    -1, & U(k,n) < 50 \% \;of \; S_n 
    \end{cases}
\end{align}

\subsection{Electric Vehicle Model} \label{sec: dynamics}
\begin{figure*}[h!]
  \includegraphics[scale=.41]{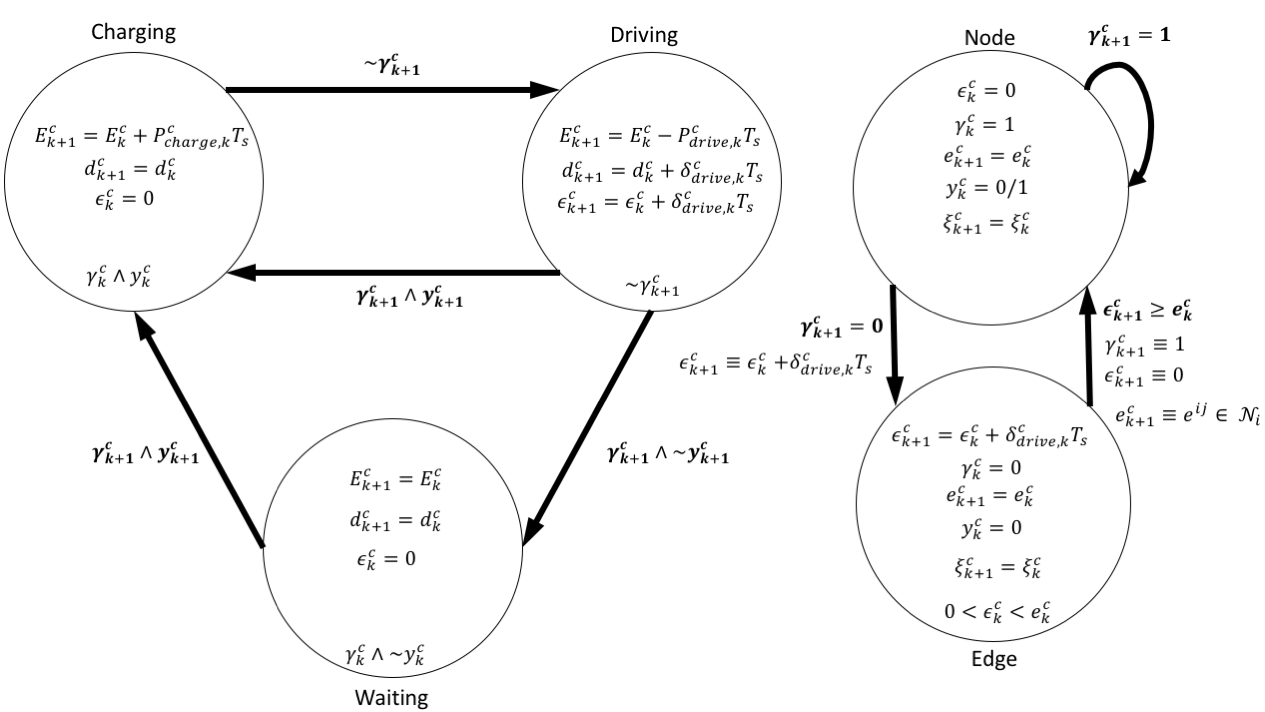}
  \caption{Finite-state machine of an EV with the extended logic finite-state machine}
  \label{fig: ev_fsm}
\end{figure*}

The EV model is defined as three-mode finite-state machine (FSM) (Fig. \ref{fig: ev_fsm}). Namely, there is a charging mode, driving mode, and a waiting mode (at the charging station). The three piecewise-continuous states are the energy in the battery, the trip distance as a scalar value, and the current edge distance traversed.

While less accurate than state of charge, the energy in the battery can be manipulated much more easily and with enough accuracy for the scheduling of trajectories. The dynamics of the battery are then a simplified version of Coulomb counting where the change in energy is a function of the power injected or consumed. Given $\frac{\partial E}{\partial t} := P(t)$ as the most basic form of Coulomb counting, let it be discretized under the one-step Euler to $\Delta E := E_{k+1} - E_{k} = P_{k}*T_{s}$, such that $E_{k+1} = E_{k} + P_{k}*T_{s}$ where $k$ denotes the discrete time and $T_{s}$ is the sampling time. This describes both the charge and discharge dynamics in this work. The charging power is a function of the energy in the battery as presented in \cite{kongetal} and the discharge power is a function of the incremental velocity, $\dot{\delta}_{k}$. Thus the switched dynamics can defined on $[\underline{E},\bar{E}]$ where \textit{Waiting} = $\beta(1)$, \textit{Charging} = $\beta(2)$, and \textit{Driving} = $\beta(3)$:

\begin{align}
    E_{k+1} &= 
    \begin{cases} 
        E_{k} + P(E_{k})T_{s}, & 
        \beta(1) \\
        E_{k}, & \beta(2) \\
        E_{k} - P_{drive}(\dot{\delta}_{k})T_{s}), &  \beta(3) \\
    \end{cases} \\
    & \sum_{i=1}^{3} \beta_{k}(i) = 1
\end{align}

The second continuous state, the discrete distance ($d_{k} \in [0,d_{max}$), is defined as a trip distance counter. $d_{max}$ is the maximum feasible distance that a car would reasonably travel on the given network. This is later useful for forming the convex hull of the continuous-valued variables. The discrete local distance $\epsilon_{k}$ has the same evolution as the trip distance when in $[0,e^{ij})$. Again, the dynamics here in continuous time are $\frac{\partial d}{\partial t} :=\dot{\delta}(t)$. In discrete-time, under the one-step Euler it becomes $d_{k+1} = d_{k} + \dot{\delta}_{k}T_{s}$. The state $d_{k}$ is primarily useful for keeping track of whether the vehicle is moving or not. Given the incremental velocity, $\dot{\delta}_{k}$, the distance counter is defined:

\begin{align}
    d_{k+1}^{c} &= 
    \begin{cases} 
        d_{k}^{c},  & \beta(1)\\
        d_{k}^{c}, &  \beta(2)\\
        d_{k}^{c} + \dot{\delta}_{k}^{c}T_{s}, & \beta(3) \\
    \end{cases}\\
    & \sum_{i=1}^{3} \beta_{k}(i) = 1
\end{align}

Similarly,  $\epsilon_{k}^{c} \in [0, {e^{ij}}^{+}]$ but with a reset at each node: 

\begin{align}
    \epsilon_{k+1}^{c} &= 
    \begin{cases} 
        0,  & \beta(1)\\
        0, &  \beta(2)\\
        \epsilon_{k}^{c} + \dot{\delta}_{k}^{c}T_{s}, & \beta(3) \\
    \end{cases}\\
    & \sum_{i=1}^{3} \beta_{k}(i) = 1
\end{align}

The guard for each of the three states is defined by boolean algebra. The extended logic is defined in Section \ref{sec: hyb_sys}. For the rest of this work, since the logic is the same for the three continuous states, the state vector will be denoted $x_{k}$.

\subsection{Costs to customers (EVs)}

The costs to the customers will be the time spent charging, the time to drive along a specified path, the degradation of the battery, and the time to wait in a queue at a full charging station.

\begin{align}
    C_{customer} := & C^{t_{charging}}(\Delta E) + C^{t_{driving}}(e^{ij}) \\
    & + C^{degrade}(\Delta E) + C^{t_{waiting}} \nonumber
\end{align}

The cost of travel time along a specific edge $e^{ij}$ was obtained from \cite{heetal}. Here $t_{e^{ij}}^0$ and $c_{e^{ij}}$ are the free flow travel time (in the absence of traffic) and capacity of link $e^{ij} \in \mathcal{A}$ respectively, which are both predetermined. Then, the travel time is a strictly increasing function of the traffic flow $v_{e^{ij}}^k$:

\begin{align}
    & C^{t_{driving}}(e^{ij}) := t_{e^{ij}} = t_{e^{ij}}^{0} * \bigg( 1 + 0.15*\bigg( \frac{v_{e^{ij}}^k}{c_{e^{ij}}}\bigg) \bigg)
\end{align}

The traffic flow $v_{e^{ij}}^k$ along a particular edge can be calculated by summing over our vector-valued exogenous input variable $\vec{\xi^c_k}$ (explained in Section \ref{sec: logic_fsm}) over all cars at any given time $k$, which gives us the total number of cars currently on that highway.

\begin{equation} \label{eq: edge_counter}
    \vec{v_{e^{ij}}^{k}} = \sum_{c=1}^{p}\xi^c_k(e^{ij})
\end{equation}

It is easiest to instead assign a quadratic cost to $C^{t_{driving}}$ that acts directly on the mode selector as a time-series vector instead of penalizing each edge. The time dependence on traffic/congestion can be separated out to be penalized independently. This will be a squared two-norm penalty on \eqref{eq: edge_counter} i.e. $\|\vec{v_{e^{ij}}^{k}}\|_2^2$ for the prediction horizon.

Both the charging time costs as well as the battery degradation model were pulled from \cite{kongetal}, where $E_i$ and $E_f$ are the
battery energies before and after the charging step.
\begin{align} \label{eq: charge_cost}
    & C^{t_{charging}}(e^{ij}) := 
    \begin{cases}
    \frac{(E_c - E_i)}{P_{max}} + \frac{1}{n}log(\frac{m-nE_{c}}{m-nE_f}), E_{i} < E_{c} \\
    \frac{1}{n} log(\frac{m-nE_{i}}{m-nE_f}), \; else
    \end{cases}
\end{align}

The charging time can be directly penalized with a quadratic cost since there is direct access to the time spent charging, $\beta(1)$. The function in \eqref{eq: charge_cost} is useful for understanding how the charging time increases with increased final state of charge, SOC.

The charging power can be derived from the charging power function in \cite{kongetal}:

\begin{align}
    & P(E) := 
    \begin{cases}
    P_{max}, \; E < E_c \\
    m - n \cdot E, \; else
    \end{cases}
\end{align}

Where m and n are constant parameters for all the charging stations, since we assume they all use identical batteries.

\begin{figure}[h!]
  \includegraphics[scale=.6]{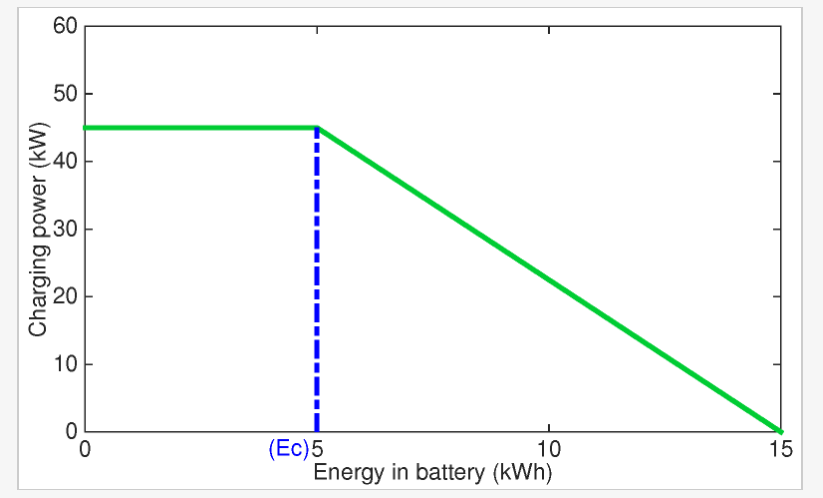}
  \caption{Charging power graph from \cite{kongetal}}
  \label{fig: charge_power}
\end{figure}

As seen in Fig. \ref{fig: charge_power} above, the rate of charging power decreases and charging duration increases as the SOC of the battery increases beyond the threshold value $E_c$. Physically, this implies that it may often be optimal to not 
charge cars to 100\% of their capacity since this would greatly slow down the charging time. Instead, the controller would have them move on to the next available station/node and let other vehicles waiting in queue charge at that station.

Once the mean charging time $\mathbb{E}(t_{charging})$ and mean charging power $\mathbb{E}(P_{ow})$ are determined using the above relations, the expected battery degradation cost can then be calculated for each EV can then be calculated as follows: 

\begin{align}
    & C_{degrade} := \bigg(a(\mathbb{E}(P_{ow}))^{2} + b*\mathbb{E}(P_{ow})+c\bigg)\mathbb{E}(t_{charging}) 
\end{align}

Where parameters a, b, and c are related to the Li-ion battery, which is the most common battery type used in EV applications.

Most of the previous literature encodes the time spent by EVs waiting at a charging outlet, as some function of the arrival rate $\lambda$ at that node and its blocking probability $\nu$, to characterize the random process.

\begin{align}
    & C_{waiting} := f \; (\lambda, \nu)
\end{align}

However, we decided to simplify matters by just penalizing the total waiting time with a quadratic cost. This was partly because neither of us had much prior experience working with queuing theory or Markov chains and also since the customer's actual waiting time can be obtained directly from our two finite state machines described below in Section \ref{sec: hyb_sys}, if necessary. This does away with the need to estimate wait-times probabilistically. 

\section{Hybrid systems theoretic formulation} \label{sec: hyb_sys}
    The EV scheduling problem is demonstrated to be piecewise continuous in the two states, $E_{k} and d_{k}$. The problem can be posed as a hybrid systems theoretic MIQP where the equality constraints are linear, the inequality constraints are affine, and the cost quadratic. Furthermore, the position of each vehicle on the highway graph can be represented as a finite-state machine in addition to the dynamics FSM. Thus, there exists two finite-state machines in this formulation. 
    
    \subsection{Framework}
    
    In accordance with the work presented in \cite{borrelli}, it is convenient to define a general framework to describe mixed discrete-time horizon automaton trajectories. The superscripts $ct$ denote continuous valued variables while $l$ denotes logical/binary variables. These superscripts are dropped outside of this section in order to simplify notation and allow for indexing cars, times, and nodes. The switched affine system, $x_{k+1}^{ct} = A^{i_{k}}x^{ct}_{k} + B^{i_{k}}\gamma^{ct}_{k}$ is defined in Section \ref{sec: dynamics}. This generally encodes piecewise-affine dynamics. The event generator, $h$, creates a binary output vector, $\delta^{e}$, that is a function of state events that satisfy some affine condition. This is considered to be the endogenous input to the other binary functions as it is an input that is decided based on the evolution of the dynamics in relation to the domains. In the framework developed in EECS C291E/ME290S, it is equivalent to a guard. Next, define the FSM, $f^{l}$, that encodes state transitions and that outputs $x^l$. Lastly, the mode selector, $\mu$, is defined to decide which discrete state, $x_{k+1}^{l}$, the finite-state machine is switching to. The general input-output relations for this work's modeling purposes are shown as follows and are not exhaustive of the work presented in \cite{borrelli}:
     
     \begin{align} \label{eq: mlda}
        \delta^{e}_{k} &= h(x_{k}^{ct},\gamma_{k}^{ct},k) \\
        x_{k+1}^{l} &= f^{l}(x^{l}_{k},\gamma^{l}_{k},\delta^{e}_{k}) \\
        i_{k} &= \mu({x^{l}_{k},\gamma^{l}_{k},\delta^{e}_{k}}) \\
        x_{k+1}^{ct} &= A^{(i_{k})}x^{ct}_{k} + B^{(i_{k})}\gamma^{ct}_{k} \\
     \end{align}
     
     To summarize these equations, the event generator is a function of the continuous valued state, $x^{ct}_{k}$, and input, $\gamma^{ct}_{k}$, where this exogenous input is zero for this work. This can also be time-variant for cases where there exist time-dependent guards. The logic FSM encodes the previously outputted value, the discrete input,$\gamma^{l}_{k}$, and the output from the event generator, $\delta^{e}_{k}$. The mode selector takes in the information from the inputs and decides what mode of the switched affine system to be in. The dynamics then propagate forward one time-step.
     
     The work can be similarly presented in the framework from EECS C291E/ME C290S; however, for formalizing the work for coding purposes with a controller, \eqref{eq: mlda} seemed to be a canonical formulation worth adopting. The analysis (Section \ref{sec: analysis}) of the resulting automaton is conducted with the tools developed in EECS C291E.
    
    \subsection{Event Generator and Guards}
    The event generator for this automaton is fairly simple. The domains of the continuous valued variables are box constraints, $x^{ct} \in [\underline{x}, \bar{x}]$ except for $\epsilon_{k} \in [0,{e^{ij}}^{+}]$, where the $+$ denotes that depending on the step value, the length of $e^{ij}$ can be overstepped in one time-step. If the domain of $x^{ct}$ is violated, then the problem is infeasible. The first element of the event generator returns $x^{l} = true$ when the upper bound of $\epsilon$ is reached. The first element of the event generator returns $false$ for $\epsilon_{k+1} \in [0,e^{ij})$.
    
    \begin{align}
        \delta^{e}_{k+1}(1) =
        \begin{cases}
        1, & \epsilon_{k} \geq e^{ij}_{k} \\
        0, & \epsilon_{k} \in [0,e^{ij})
        \end{cases}
    \end{align}
    
    The second element of the output of the event generator indicates whether the vehicle is moving on an edge, or has arrived at a node:
    
    \begin{align}
        \delta^{e}_{k+1}(2) =
        \begin{cases}
        1, & \epsilon_{k} \in (0,e^{ij}) \\
        0, & \epsilon_{k} \notin (0,e^{ij})
        \end{cases}
    \end{align}
    
    \subsection{Logic FSM} \label{sec: logic_fsm}
    The logic FSM (boolean output) keeps track of what edge a car is currently on. In particular it encodes the event generator as well as the exogenous input variable, $\gamma_{k}^{l}$. This variable decides whether the vehicle is at a node or on an edge. It is bounded under certain conditions:
    
    \begin{align}
        \gamma_{k+1}^{l} = 
        \begin{cases}
        1, & \delta_{k}^{e}(1) \\
        0/1, & \neg \delta_{k}^{e}(1) \\
        0, & \delta_{k}^{e}(2) \\
        \end{cases}
    \end{align}
    
    This indicates that the exogenous input is $true$ when the event generator is also $true$ and can be $\{0,1\}$ when the event generator returns $false$ since this physically means that the car can stay or leave a node when it wants to. The input is further restricted when it is on an edge, which is the third case; it says that it has to be zero when it is traversing an edge.
    
    The value of the second input variable, $y_{k}^{l}$, indicates whether a car is charging or not:
    
    \begin{align}
        y_{k+1}^{l} &= 
        \begin{cases}
        0, & \neg \gamma_{k+1}^{l} \\
        0, & \gamma_{k+1}^{l} \land \sum_{c} y_{k}^{ct}(c) = cap^{n}_{k} \land y_{k} = 0\\
        0/1, & \gamma_{k+1}^{l} \land \sum_{c} y_{k}^{ct}(c) = cap^{n}_{k} \land y_{k} = 1\\
        0/1, &  \gamma_{k+1}^{l} \land \sum_{c} y_{k}^{ct}(c) < cap^{n}_{k}\\
        \end{cases} \\
        & \sum_{c=1}^{p} y_{k+1}^{ct}(c) \leq cap^{n}_{k+1} \label{eq: stat_con}
    \end{align}
    
    If the car is at a node, indicated by $\gamma_{k+1}^{l}$, then the car \textit{may} charge. If it is not at a node, then it \textit{cannot} charge. An additional charging station capacity constraint, denoted $cap^{n}_{k}$ ($\in \mathbb{Z}$) is included as stations can only provide service to a certain number of EVs at a time. For continuity of charging, the condition $y_{k} = 1$ is included to allow for cars to keep charging at the next time step if they already were in the previous time step, or $y_{k} = 0$ prevents cars from charging if they hadn't been previously and a station is at capacity. This is where the time indexing becomes muddled. The logic FSM is generally defined to output one-step ahead. However, since the first input affects the second input in a disjunctive condition, it should presumably still be labeled as $k+1$. To check this, the event generator at $k$ could be introduced directly into the condition for the second input. Thus, the time index for this is correct. The time index for the capacity condition is also non-obvious. The way it is written indicates that if the station is currently at capacity, then new cars can't start charging until one time step has passed during which the station is not at capacity. While this is not optimal for large time steps, it creates an appropriate condition. \eqref{eq: stat_con} makes sure that numerous EVs don't start charging in the next time step and exceed the station's capacity constraints.
    
    The third, vector-valued exogenous input variable is $\xi_{k}^{l}$. This encodes which edge a vehicle is currently on. Specifically, $\xi_{k}^{l} \in \{0,1\}^{N^{2}}$, or can be represented in a $NxN$ matrix corresponding to whether an edge is selected, since there are a maximum of $N^2$ edges in a graph with $N$ nodes. For example, in the vector form, if $\xi_{k}^{l}(5)$ is selected in a five-node graph, then it indicates the vehicle is traveling from node 1 to node 5. This allows for the next edge to be selected each time a car reaches a node. Under the FSM framework with $h$ acting as the index in $[1,N^2]$:
    
     \begin{align}
        \xi_{k+1}^{l}(h) &= 
        \begin{cases}
        \xi_{k}^{l}(h), & \neg \delta_{k}^{e} \\
        0/1, & \delta_{k}^{e} \land h \in \mathcal{N}_{\xi_{k+1}^{l}} \\
        0, & \delta_{k}^{e} \land h \notin \mathcal{N}_{\xi_{k+1}^{l}}
        \end{cases} \\
        \sum_{h=1}^{N^2} \xi_{k}^{l}(h) &= 1
    \end{align}
    
    The notation for the latter two cases is poor but essentially indicate that $\xi_{k+1}^{l}$ can only be true if that edge is connected to the current node. Here the constraints on the exogenous inputs under specific constraints were shown but the unconstrained conditions (as it is not deterministic) allow for a solver to choose values that minimize the objective function. The output of the FSM is then a binary vector $x_{k+1}^{l} := [\gamma_{k+1}^{l},y_{k}^{l},(\xi_{k}^{l})^{T}]^{T}$.
    
    \subsection{Mode Selector}
    
    The discrete mode is then selected as a function of the results from the event generator, the exogenous input, and the logic FSM. Let \textit{Waiting} = $\beta(1)$, \textit{Charging} = $\beta(2)$, and \textit{Driving} = $\beta(3)$ be binary variables in this illustration of the mode selection:

    \begin{align}
        \mu_{k+1}^{l} &= 
        \begin{cases}
            \beta(1), & \gamma_{k}^{l} \land y_{k}^{l} \\
            \beta(2), & \gamma_{k}^{l} \land \neg y_{k}^{l} \\
            \beta(3), & \neg \gamma_{k}^{l}
        \end{cases} \\
         \sum_{i=1}^{3} \beta_{k}(i) &= 1
    \end{align}

    The utility of $\xi_{k}^{l}$ becomes apparent now as the evolution of the current edge must be considered. If a node $j \in \mathcal{N}_{i}$, then $\xi_{k}^{l}(h)$ \textit{can} be positive. The algebra for this relation is inconvenient and non-intuitive (Section \ref{sec: appendix}) but the relation is as follows:
    
    \begin{align} \label{eq: edge_sel}
        e_{k+1} &= 
        \begin{cases}
            e_{k}, & \neg \delta_{k}^{e} \\
            \begin{split}
            \xi_{k}^{l}(h)e^{ij}, & \delta_{k}^{e} \land j \in \mathcal{N}_{i}, \\ & 
            \xi_{k}^{l}(h)e^{ij} > 0,  \forall h \in \{1,...,N^2\} \\
            \end{split}
        \end{cases}
    \end{align}
    
    The incremental velocity mentioned in Section \ref{sec: dynamics} is defined as a non-constant increment as it decreases with congestion on an edge and is highest in free-flow conditions. In general it is of the form $\dot{\delta}_{k} := \frac{e^{ij}}{t_{ij}}$ where $t_{ij}$ is the time to traverse an edge. For future work this can be exploited but for now it is left as constant.
    
    Thus, the mode for the switched dynamics, as well as the next edge can be chosen. This is also illustrated in the left-hand FSM of Fig. \ref{fig: ev_fsm}.
    
\section{Analysis from a Hybrid Systems Perspective} \label{sec: analysis} 

The EV and highway network models described in the previous sections incorporates several elements of hybrid systems, according to the framework developed in the ME 290S/EE291E class. These can be thought of as both 1) continuous, embedded systems (in distance travelled and battery energy level) controlled by discrete logic and 2) multi-agent subsystems, with many subsystems (EVs) interacting with one another. 


We can explicitly define the following entities for the hybrid automaton $\mathcal{H}(Q,X,Init,f,Dom,R)$ of the EV model described above: 
\\
\begin{itemize} 
    \item Discrete states of the EV $q \in Q$ where $Q = \{waiting \; \beta(1), charging \; \beta(2),
    driving \; \beta(3)\}$ is a finite collection of states possible for the two FSMs shown in Fig. \ref{fig: ev_fsm}. 
    \\
    \item Continuous states $X = \{d_k^c, E_k^c,\epsilon_{k}^{c}\} \subseteq \mathbb{R}^3_{+++}$ for the distance traveled from the source $d_k^c$ and energy level of the battery $E_k^c$ or equivalently state of charge $SOC = \frac{E_k^c}{E^c_{max}}$, and $\epsilon_{k}^{c}$ is the counter for distance traveled along a specific edge.
    \\
    \item Discrete inputs $\Sigma = \{ \gamma_k^c, y_k^c, \vec{\xi}_k^c$ \}, where $\gamma_k^c \in \{0,1\}$ switches the EV between nodes and edges, $y_k^c \in \{0,1\}$ switches it between charging and waiting, and $\xi_k^c \in \{0,1\}^{N^2}$ chooses between edges at a node.
    \\
    \item Continuous inputs $V = \o$
    \\
    \item Initial state $Init \subset Q \; x \; X$ where the car always starts off by waiting at a source node $v_{0,i}$ at a certain distance $d_k^c$ from the origin, with an initial energy level $E_k^c \ge 0$ and $\epsilon_k^c = 0$
    \\
    \item Continuous dynamics $f$: Both the battery energy level and distance traveled vary continuously and can be represented using continuous state, continuous time differential equations as a vector field. However, one step Euler discretizations were used for this digital control system  instead, using the charging/discharging powers and the incremental velocity $\dot{\delta}_k$ respectively.
    \\
    \item The event generator $\delta^e_k$  is the guard that switches the EV between edges and nodes in the first FSM, and thus keeps the EV automaton inside the domain (e.g. this ensures that the EV never violates the prescribed domain by running out of charge while traveling along an edge).
    \\
    \item A transition relation or reset map $R: Q \; X \; \Sigma \mapsto 2^Q$ that describes the switching logic of the finite automaton. In our automaton, this is the mode selector $\mu$ that switches between the discrete states of charging, waiting, and driving states in the second FSM. \\
    
    \item Domain for the endogenous inputs, \\
    $Dom = (\beta(1), \{d_k^c \ge 0, E_k^c \ge 0, \epsilon_k^c = 0\}) \\ \cup \; (\beta(2), \{d_k^c \ge 0, E_k^c \ge 0, \epsilon_k^c = 0\}) \\ \cup \; (\beta(3), \{d_k^c \ge 0, E_k^c > 0, \epsilon_k^c \ge 0\}) $ \\
    where $\cup$ denotes the union of sets
    \\
\end{itemize}

We attempted to analyze this system from a reachability standpoint but it is computationally challenging and time intensive to explicitly calculate the feasible trajectories/paths as well as all their possible permutations given a particular starting node and initial energy level. However, we can clearly see that all nodes and states in our EV network's domain are reachable with finite executions, even without explicitly computing the reachable sets. This is because we can assume that any realistic highway road network will most likely be a connected graph i.e. there exists at least one path that you can take between every pair of vertices. Thus given a particular starting node, an EV can reach any other node in finite time and without running out of charge.

Furthermore, we can also argue that our automaton is domain preserving (i.e. the domain of possible discrete states (modes) and continuous states (state of charge and distance traveled $\in \mathbb{R_+}$) is an invariant set). Since our initial set of states is in the domain, and all the reachable states in the network from these initial states is in the domain, the hybrid automaton is domain
preserving i.e 
\begin{center}
    $Init \subseteq Dom \implies Reach \subseteq Q \; X \; Dom.$
\end{center}

Similar to what we learned in the class, we can then proceed to also define hybrid time sets $\tau$ and trajectories/executions $(\tau, q, x)$ over which our EV system (automaton) is allowed to evolve. For the EV system, we can envision hybrid trajectories with either finite, finite-open or Zeno hybrid time sets. Since $R(q,x) = \mu_k^c \ne \phi \; \forall$ states $(q,x) \in Dom$, we can conclude that $\mathcal{H}$ is non-blocking. This means that for all reachable states for which continuous evolution (in distance or energy) is impossible, a discrete transition (i.e. switching to a different mode) is possible. Of course, this is only true when the edges are short enough such that a car can traverse an edge without running out of energy and the horizon is long enough such that cars can wait to charge before driving further.

The automaton is also non-deterministic since 1) continuous evolution in $d_k^c$ and $E_k^c$ is possible even when discrete transitions between the modes $\{\beta(1),\beta(2),\beta(3) \}$ occur and 2) each discrete transition between any two of these modes could lead to multiple different destinations, implying that $|R(q,x)| \nleq 1$ in general. This makes intuitive sense since given a particular starting node (initial condition), the EV could reach several different destinations and thus there are a whole family of state trajectories (solutions) possible. There could also be multiple ways to travel between the same start and end nodes while meeting all the constraints and minimizing our objective (especially for larger networks than our 5-node example).

Thus, the non-blocking property implies the existence of finite and infinite executions for all initial states i.e. local existence of solutions is guaranteed. Non-determinism implies that even though all executions $(\tau, q, x)$ can be extended to infinite executions, these are not necessarily unique. 

By construction, the EV system defined above is also non-Zeno since our model only allows one discrete transition per time step. Thus, $\mathcal{H}$ can never accept Zeno executions since it's impossible to complete infinite discrete transitions within finite time.

\section{Optimization Problem}

The modeling framework proposed in the previous section is for each car. In order to account for each car, the conditions hold for each vehicle on the network. Interactions between the cars are only considered insofar as they affect the incremental velocity, $\dot{\delta}_{k}$, and the availability of a charger at a station.

The idea behind this formulation is that it will be run as a receding horizon control (RHC) problem. The prediction horizon, Hp, will be computed, applied, and then calculated again at the next sampling time in the control horizon $H_p$.

\begin{align}
    & \min_{\{y,\gamma  \in \{0,1\}^{H_p}, \xi \in \{0,1\}^{H_p,N^{2}}\}^{p}} C_{total} \\
    & \text{ s.t.  } \nonumber \\ 
    & C_{total} = C_{stations} + C_{customers} + C_{electricity} \\
    & \eqref{eq: mlda} \nonumber \\
    & \underline{x} \leq x_{k} \leq \bar{x}, \qquad \forall c, \forall k \in \{0,1,...H_p\} \\
    &\epsilon_{0} = 0, \quad \forall c, \forall k \in  \{0,1,...H_p\} \\
    &\epsilon_{H_p} = 0, \quad \forall c, \forall k \in \{0,1,...H_p\} \\
    &\xi_{0}(\nu_0) =  true \quad \forall c \\
    &\xi_{H_p}(\nu_{H_p}) = true \quad \forall c \\
     \nonumber
\end{align}  

The constraints on the problem are the mixed logic dynamics defined in Section \ref{sec: dynamics} and \ref{sec: hyb_sys}. The box constraints here are reiterated for clarity, and the initial and terminal constraints are added. The counter variable, $\epsilon_{k}$, is zero at the start and the end. The starting and ending nodes are dictated by $\xi_{k}$; the actual indexing on $\xi$ is non-trivial as it relates the coming-from and going-to a node in index arrays. It is included in the Appendix (Section \ref{sec: appendix}) for brevity here but are fully-defined in how a node is selected.

\subsection{Disciplined formulation}
In order to convert this MIQP into a problem for a numerical solver, it must be reformulated into a problem that has "good" numerical properties and uses the tools available in disciplined convex programming.

The canonical way of expressing piecewise functions is via the big-M relaxation \cite{borrelli}, \cite{johan}. The problem,

\begin{align} \label{eq: switched}
    x_{k+1} &=
    \begin{cases}
    A^{1}x_{k} + B^{1}u_{k}, & \neg \delta_{k}\\ 
    A^{2}x_{k} + B^{2}u_{k}, & \delta_{k}\\
    \end{cases}
\end{align}

can be formulated to be,

\begin{align}
    x_{k+1}
    \begin{cases}
    (m_{1}-M_{2})\delta_{k} + x_{k+1} \leq  A^{1}x_{k} + B^{1}u_{k} \\
    (m_{2}-M_{1})\delta_{k} + x_{k+1} \leq  A^{1}x_{k} + B^{1}u_{k}\\
    (m_{2}-M_{1})(1-\delta_{k}) + x_{k+1} \leq  A^{2}x_{k} + B^{2}u_{k} \\
    (m_{1}-M_{2})(1-\delta_{k}) + x_{k+1} \leq  A^{2}x_{k} + B^{2}u_{k}\\
    \end{cases}
\end{align}

The four inequality constraints represent the two equality constraints where the first statement of \eqref{eq: switched} is tight when the logical, $\delta_k$ is $false$ in the first two constraints and the second statement of \eqref{eq: switched} is tight when $\delta_k$ is $true$. The point of interest is when the bound is not tight. In this case the values of $m_{i},M_{i}$ are important as they define the box constraints on the continuous variable $x_{k} \in \mathbb{R}$. If the bounds are arbitrarily small, (for $m_{i}$ it means $|m_{i}|$ is large). Then the domain for $x_{k}$ is very large. This leads to bad numerics for solvers. For bounds that are on the order of the actual solutions bounds, the solve time can be longer than necessary due to extra branching \cite{johan} in branch-and-bound solvers.

Thus, "good" means that the convex hull of the piecewise-polytopes is found. This is generally non-trivial as the feasible set at a specific timestep is a subset of the box constraints that define the domain of the states and the inputs for all time. Fortunately, the modeling languages YALMIP provides a tool that allows for direct implications to be made. YALMIP uses the function \textit{implies(a,b)} for logic programming to imply that if \textit{a} is $true$, then \textit{b} is too. If \textit{a} is $false$, then \textit{b} may or may not be $false$ as well.

While \textit{implies()} is convenient in clarity of expressions, it does have existing bugs. For instance, through talking with the developer of YALMIP, it was discovered that the operator \textit{not()} was not overloaded while all the other MATLAB logical expressions were. Errors such as this while using a somewhat black-box tool (based on big-M but with many numerical tools adjusting it) made debugging the program more difficult.

The problem was coded in both YALMIP using \textit{implies()} and purely with the big-M formulation. While including more binary variables in both cases is cumbersome and intuitively more complicated for a solver, it actually creates cleaner big-M relaxations in both cases. The reason for this is that it creates disjunctive behavior between conditional cases involving multiple continuous-valued variables. For example, the trivial implication "if $x > 5$, then $y < 3$" where y is the variable that should be constrained can result in the value of x being modified so that the condition is true. A disjunctive equivalent would then include $\delta \in \{0,1\}$ so that "if $x > 5$, then $\delta$. If $\delta$, then $y < 3$".

\section{Process and Preliminary Results}

\subsection{Process}
Various reformulations of the problem have been undertaken over the past two months. Originally, the problem used a model that was more consistent with previous work done \cite{heetal},\cite{kongetal}. In order to keep it in the same space as more hybrid system theoretic approaches, a first draft of the current model was proposed. 

The first draft incorporated the basic dynamics of keeping track of trip distance and energy in the battery. In order to reduce the order of computational complexity, it was thought that it would be best to pre-compute the $n$ shortest total travel time trajectories, for example $n=3$, under fully available charging stations and zero traffic congestion (i.e. free-flow). While this would seem to reduce the solution space when using a branch-and-bound solver as the number of "branching" variables is reduced, the scaling of this was very poor because every combination of each car's $n$ feasible trajectories would need to be computed with every other car's $n$ feasible trajectories.

The next step was to allow for the model to switch edges itself. While this would increase the complexity of each problem being solved, it removed all of the pre-computation. Since moving to this scheme, the basic model has been modified multiple times. The main modifications have been the guard and reset relations. The guards for the problem were originally defined based on the relation of the counter variable $\epsilon$ to the current edge length in both transitions to and from a node. This often led to blocking cases where once in the driving state, the vehicle could not stop again to charge.

Throughout this process, it was difficult to understand how relaxing a constraint would affect the feasible set. Generally, visualizing N-step invariant sets is straightforward with time-invariant dynamics that are continuous-valued functions. With mixed-integer programs it becomes more difficult. YALMIP has convenient plotting functions: the constraints can be formed into n-dimensional polyhedra and projected down onto arbitrary lower dimensions. Unfortunately, the computation time for using this for mixed-integer constraints is prohibitive due to the large number of binary variables.

One of the key areas that is misunderstood in this formulation is the evolution of the logical operators as defined in Section \ref{sec: logic_fsm}. Somewhat counter-intuitively, there are numerous cases where "if $z_{k+1}$, then $w_{k+1}$", with $z_{k+1}, w_{k+1} \in \{0,1\}$, is different (i.e. feasible/infeasible) than "if $z_{k}$, then $w_{k}$" where the entire disjunctive logic's time indices is changed appropriately.

\subsection{Preliminary Results}

\subsubsection{Improper constraints}
The current version of the code exhibits somewhat reasonable behavior when constrained under the switched dynamics and logic defined in Section \ref{sec: model} but there exists faulty edge conditions due to unknown errors in the discrete time logic.

\begin{figure}[h!]
  \includegraphics[scale=.24]{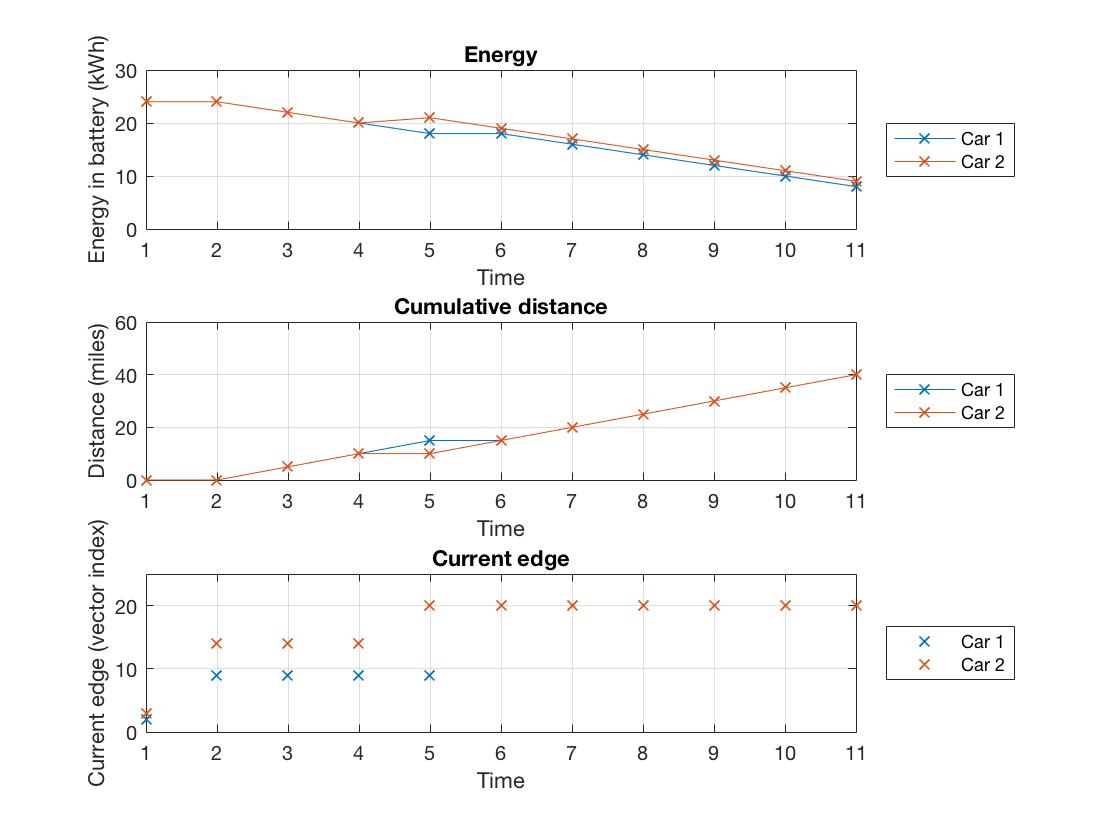}
  \caption{Example trajectories of two cars on a five-node network. The current edge is characterized as the vector shown in Section \ref{sec: appendix}.}
  \label{fig: car_traj}
\end{figure}

\begin{figure}[h!]
    \centering
  \includegraphics[scale=.23]{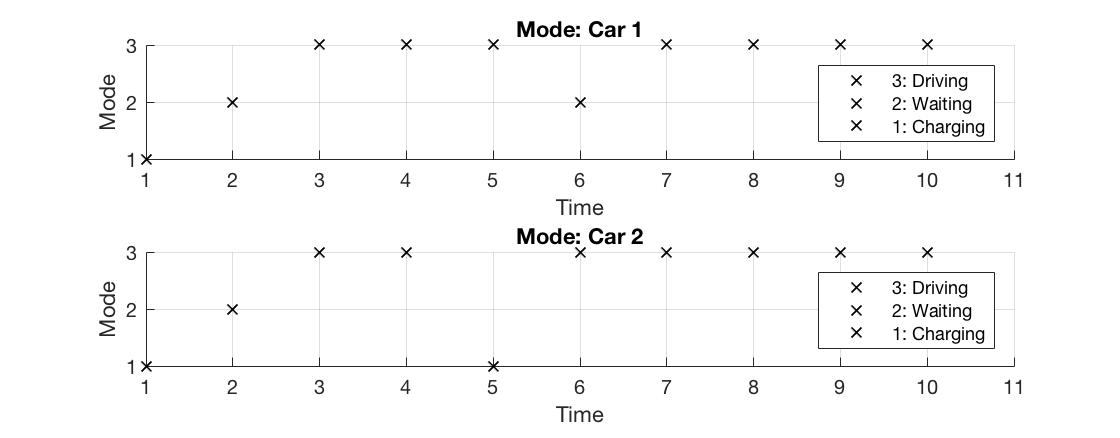}
  \caption{The mode of the two cars (Fig \ref{fig: car_traj}) as a time-series is shown with the three discrete modes of the EV}
  \label{fig: modes}
\end{figure}

In particular, Fig. \ref{fig: car_traj} shows the trajectory of a vehicle under all of the constraints save for the station capacity. The objective here is to simply maximize the distance traveled via a quadratic penalty on the difference between the terminal distance and some arbitrarily large distance. This is the test cost function to exploit the modes and the dynamics. In the example, it can be seen from Fig. \ref{fig: modes} that the first car does not charge despite reaching a switching condition (a node) where it can charge under the relaxed constraints. Instead, it waits at a node. This reduces the cumulative distance traveled and thus exposes an issue in the logic defined above. The second car, however, is able to charge, and does so for one time-step. Ostensibly, neither should be waiting in the beginning but do so. This is thought to be the manifestation of issues with initialization and time indexing errors at this point in the project.

\begin{figure}[h!]
  \includegraphics[scale=.21]{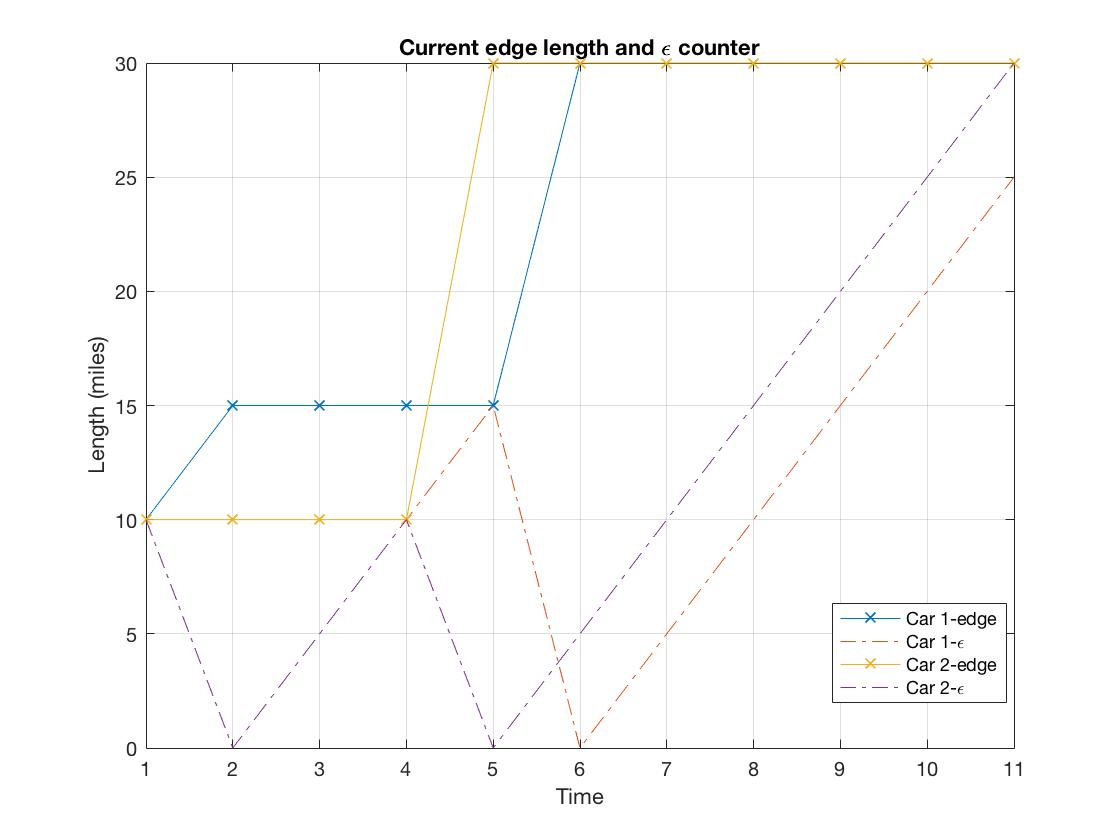}
  \caption{The local edge length is kept track of the and the local distance is encoded by the counter state denoted as $\epsilon$.}
  \label{fig: traj_counter}
\end{figure}

The local distance counter, $\epsilon$, is shown in Fig. \ref{fig: traj_counter} to be increasing over $H_p= 10$ and resets when a node is reached. The counter logic is thus illustrated to function well. The fourth quadrant indicates that the current edge changes. This is indicated in terms of the vector-value of the indicator function $\xi$. The edges take values defined by the set of edges, $\mathcal{E}$, then and can change at a node.

The terminal condition is numerically poor in this example where the end-node constraint cannot be written explicitly due to the construction focused on edges and adjacency instead of nodes. This results in the condition: \textit{At \text{Hp}, the car must be within $\Delta$ of the final node, \text{j}, (either pointing to \text{j} (i.e. ij) or from \text{j} (i.e. ji))}. The formulation is non-intuitive making it difficult to debug. In this example the cars should point from the end destination node 4. This means that it can take the value (4,9,14,19,24) insofar as that edge is connected by a node that is incident with the previous edge. The resultant edge is in fact labeled 20, which corresponds to node 5 pointing at node 4. This outcome occurs because the terminal constraint on $\epsilon$ was relaxed to maintain feasibility. The terminal constraint here needs to be reformulated to be more intuitive.

\subsubsection{Computational complexity- scaling}
\begin{figure}[h!]
  \includegraphics[scale=.24]{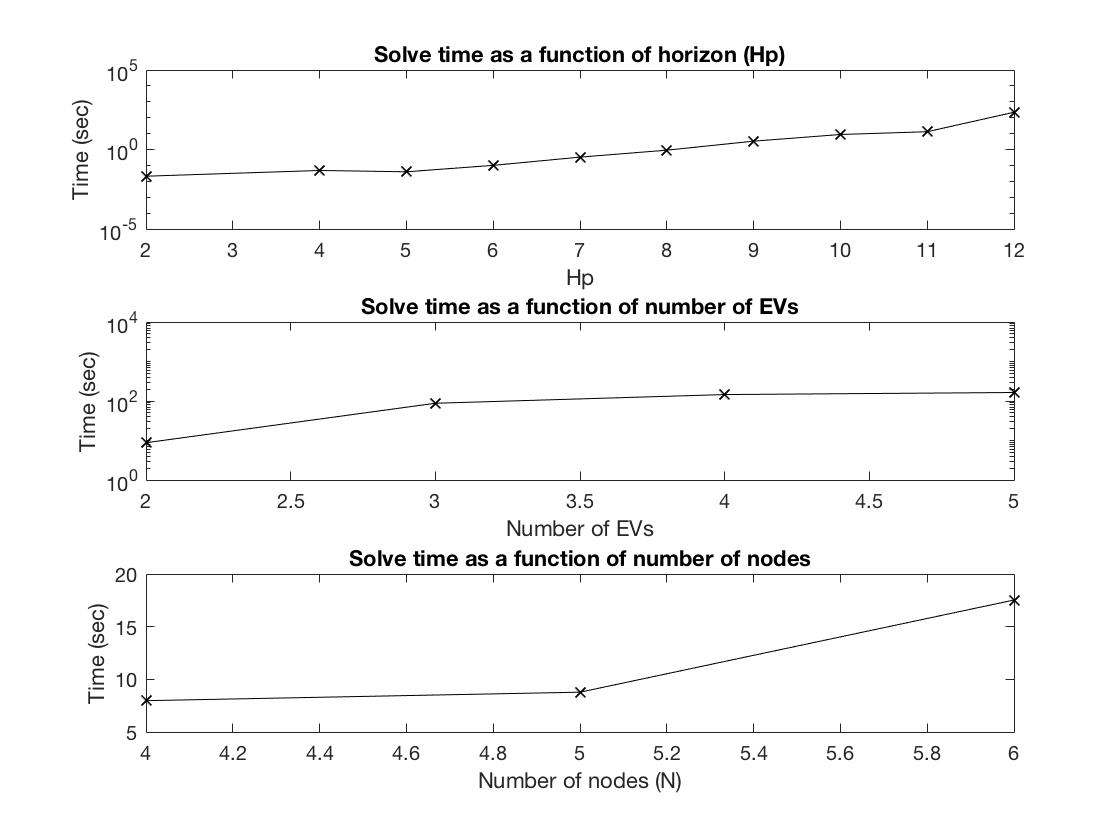}
  \caption{The scaling complexity of the problem is shown here. The primary influence is the prediction horizon length}
  \label{fig: comp_complexity}
\end{figure}

Another issue is that the problem scales poorly. The parameters that change with the problem parameters are $H_p$, $C$, and $N$. Fig. \ref{fig: comp_complexity} indicates that the time to solve the problem increases dramatically with increased prediction horizon, Hp, over $H_{p} = 10$. Increasing the number of EVs has diminished effects at some threshold and is thus not considered to be as much of a computational barrier. The number of nodes in the network does not seem to have a drastic effect on computation but larger networks have not been tested.

The computational issue lies in the complexity underlying the prediction horizon. In particular, there is a trade-off between the time-step size and the horizon length. With larger time-steps, the precision of the model begins to falter and cars can overshoot nodes (i.e. $\epsilon$ can attain a value much greater than its current edge length in one time-step). This results in inaccurate characterization of the trip distance of a vehicle. Smaller time-steps improve this. However, in order to reach a terminal node from a defined initial node, the time horizon must be sufficiently large. The requirements for these two parameters are inversely proportional. As a result of this complexity and desired small time-step, the current algorithm (i.e. the exact solution provided by Gurobi--a branch-and-bound solver) is not a good option.

\section{Next steps}

Moving forward, we aim to polish and fix MATLAB code bugs in our most recent approach for programming the logic FSM and counter variables. In particular, the definitions for the initial and terminal conditions need to be refined or reformulated. This will allow the optimal trajectory to be computed correctly given starting and ending nodes on the small example network. 

However, we recognize that our current solution does not scale well with larger networks as we increase the time horizon and number of participating agents (stations and EVs) greatly. Considering the relatively high-level nature of this scheduling formulation, it may make sense to explore heuristic approaches that scale very well and compromise the optimality of the solution.
Further time should be spent visualizing the feasible sets and understand how they vary with time. Understanding the nuances of the variant sets will enable the logic to be smoothed out and the time indices to be coded with more rigorous reasoning.

The last step in mind is to simulate this using data from real highway networks that have high EV penetration, multiple charging stations that have finite capacity, and many, perhaps directed, edges. We will try to use data sets that have already been used by \cite{heetal} and \cite{kongetal} in the past. 

\section{Conclusion}

This work demonstrated the formulation of an EV scheduling model that is hybrid systems theoretic. The model is represented using two frameworks, one that is more canonical from a control standpoint, and one that is more rigorous when considering desirable properties of the model. The simulations indicate that the model has logical, or at least indexing, errors. The simulations illustrate the working principle of a switched dynamical system with extended decision-making logic.

Cost functions accounting for the aggregate cost to all players were introduced, many drawn from recent works in the field and cited as such. The utility of creating such a cost function is yet to be realized operationally.

The immediate next steps for this work are laid out so that the model can be  implemented in simulations using data from real highway networks.

\section{Acknowledgements} 

We would like to thank Prof. Claire Tomlin and Forrest Laine for their informative in-class lectures and discussions related to modeling hybrid systems, as well as for offering advice and help during office hours. We also thank Prof. Francesco Borrelli since we drew a lot of inspiration for our latest hybrid systems theoretic approach from his book on Model Predictive Control. Finally, we would like to thank Johan Lofberg for offering some assistance in working through existing issues with \textit{implies()} in YALMIP.

\section{Appendix}\label{sec: appendix}
\subsection{Indexing with respect to edges and nodes}
The constraints are built iteratively. In particular, in the YALMIP modeling language with the "constraints" list recursively defined:

\begin{lstlisting}
% create the switching conditions for edges
for i = 1:n_nodes
    % Consider the cases where a switch must be made due to reaching a node
    % There are n of these cases since this can occur at each node in a
    % network.
    next = (i-1)*n_nodes;
    xi_idx = i:n_nodes:n_nodes^2;
    %             disp([num2str(i) ' | ' num2str(xi_idx) ' | ' num2str(next+1:next+n_nodes)]);
    
    constraints = [constraints,...
        implies(state_preserve(c,k+1) + sum(xi{c}(xi_idx,k))== 2,...
        current_edge(c,k+1) == xi{c}(next+1:next+n_nodes,k+1)'*E(:,i)),...
        ];
    
end
\end{lstlisting}

\end{document}